\documentclass[pdflatex,sn-nature]{sn-jnl}

\usepackage{graphicx}%
\usepackage{multirow}%

\usepackage{amsmath,amssymb,amsfonts}%
\usepackage{amsthm}%
\usepackage[title]{appendix}%
\usepackage{xcolor}%
\usepackage{textcomp}%
\usepackage{manyfoot}%
\usepackage{booktabs}%
\usepackage{algorithm}%
\usepackage{algorithmicx}%
\usepackage{algpseudocode}%
\usepackage{listings}%

\DeclareMathOperator{\Tr}{Tr}

\begin{document}

\title[Article Title]{Precision thermodynamics of the strongly interacting Fermi gas in two dimensions}

\author[1]{S. Ramachandran}
\author[2]{S. Jensen}
\author[1]{Y. Alhassid}

\affil[1]{Center for Theoretical Physics, Sloane Physics Laboratory, Yale University, New Haven, Connecticut 06520, USA}
\affil[2]{Department of Physics and Astronomy, University of Louisville, Louisville, Kentucky 40241, USA}

\abstract{
\unboldmath
The two-species cold atomic Fermi gas with attractive short-range interactions in two spatial dimensions undergoes a Bardeen-Cooper-Schrieffer (BCS) to a Bose-Einstein Condensate (BEC) crossover as a function of $\ln (k_F a)$, where $a$ is the scattering length. However, the nature of this crossover in the strong coupling regime $\ln(k_F a) \sim 1$ remains poorly understood. In this work we use canonical-ensemble auxiliary-field quantum Monte Carlo methods on discrete lattices to calculate several thermodynamical quantities in the strongly interacting regime, and eliminate systematic errors by extrapolating to continuous time and taking the continuum limit.  In particular, we present results for the condensate fraction, spin susceptibility, contact, energy equation of state, and the free energy staggering gap.  We identify signatures of a pseudogap regime, in which pairing correlations survive above the critical temperature for superfluidity, in the spin susceptibility and in the free energy staggering gap. These results can be used as a benchmark for future experiments.
}

\maketitle

\section{Introduction}
In recent decades, ultracold quantum gases have emerged as a powerful platform for exploring foundational questions in quantum many-body physics. Advances in experimental techniques have enabled controlled realizations of quasi-two-dimensional (2D) systems. In particular, the two-dimensional spin-1/2 Fermi gas with attractive short-range interactions has been realized in the strongly correlated regime. This capability enables precise studies of strongly correlated 2D superfluidity and motivates the development of new theoretical tools to quantitatively describe the underlying phenomena.

The ground state this system undergoes a crossover as the coupling parameter $\ln(k_F a)$ is varied, where $k_F$ is the Fermi momentum, and $a$ is the $s$-wave scattering length. In the limit $\ln(k_F a) \rightarrow +\infty$, the ground state is a Bardeen-Cooper-Schrieffer (BCS) fermionic superfluid. In the limit $\ln(k_F a) \rightarrow -\infty$, the ground state is a Bose-Einstein condensate (BEC) of dimer molecules, bosonic bound states of two fermions. Between these limits lies the crossover regime, with strong correlations for $\ln(k_F a) \sim 1$. The nature of this regime, characterized by enhanced fluctuations, remains poorly understood, and the absence of a small parameter poses a significant challenge for theorists.

The 2D BCS-BEC crossover shares some similarities with its three-dimensional counterpart, a system of longstanding theoretical and experimental interest. However, there are key differences that render the 2D case unique. The superfluid phase transition in two dimensions belongs to the Berezinskii-Kosterlitz-Thouless (BKT) universality class. Moreover, unlike in three dimensions, a two-body state exists in two dimensions for arbitrarily weak attractive interaction strength.

A central open problem in the 2D BCS-BEC crossover is the extent and nature of a possible pseudogap regime, in which pairing correlations persist above the BKT superfluid transition temperature $T_c$. While pseudogap phenomena have been extensively studied in three dimensions~\cite{Jensen2020Pairing,Jensen2024}, the stronger fluctuations and universal presence of a two-body bound state make the 2D case particularly interesting. Understanding how high-temperature pairing correlations emerge beyond the two-body effects remains an outstanding theoretical challenge. 

There have been many experimental~\cite{Frolich2011,Vogt2012,Murthy2015,Ries2015,Boettcher2016,Toniolo2017,Luciuk2017,Hueck2018,Murthy2018,Feld2011} and theoretical~\cite{Watanabe2013,Matsumoto2014,Bauer2014,Anderson2015,Marsiglio2015,Shi2015,Galea2016,Vitali2017,Madeira2017,Schonenberg2017,Mulkerin2018,Hu2019,Wu2020,Pascucci2020,Zhao2020,Zielinski2020,Mulkerin2020A,Mulkerin2020B,Wang2020,He2022} studies of the 2D BCS-BEC crossover. The spectroscopic measurements of Ref.~\cite{Murthy2018} presented an energy pairing gap as a function of interaction strength at high temperatures ($T\sim0.5\, T_F$). These results indicated a high-temperature pairing regime, with a pairing gap well above $T_c$, consistent with a strong pseudogap regime with strong correlations. Additionally, at $\ln(k_F a)\sim1.0$, the energy gap was found to be significantly larger than the two-body binding energy, indicating the presence of a ``many-body pairing regime'' and non-trivial pseudogap phenomena. 

The current work is a follow up paper to a recently published letter~\cite{Ramachandran2024}. Here we present further auxiliary-field quantum Monte Carlo (AFMC) results for the thermodynamic properties of the system, exploring the condensate fraction, spin susceptibility, energy equation of state, Tan's contact, and the free energy staggering gap. We compare these results to experimental results when available. Finally, we identify key directions of further research for the 2D Fermi gas in its strongly correlated regime.

\section{Methods}
To achieve a controlled theoretical study of the 2D BCS-BEC crossover, we model the contact interaction using a continuum Hamiltonian,
\begin{equation}\label{continuumhamiltonian}
\hat{H} = \sum_{s_z} \int d^2 \mathbf{r} \ \hat{\psi}_{s_z} ^\dagger (\mathbf{r}) \left ( -\frac{\hbar^2 \nabla^2 _{\mathbf{r}}}{2m} \right ) \hat{\psi}_{s_z} (\mathbf{r}) + V_0 \int d^2 \mathbf{r} \ \hat{\psi}_{\uparrow} ^\dagger (\mathbf{r}) \hat{\psi}_{\downarrow} ^\dagger (\mathbf{r}) \hat{\psi}_{\downarrow} (\mathbf{r}) \hat{\psi}_{\uparrow} (\mathbf{r}) \;,
\end{equation}
where $\hat{\psi}$ are the fermionic field operators, and we assume a gas with two spin species, $s_z = \uparrow,\downarrow$. To solve this system numerically with AFMC, we discretize the continuum model with a finite lattice with periodic boundary conditions in both dimensions,
\begin{equation}\label{latticehamiltonian}
\hat{H} = \sum_{\mathbf{k}, s_z} \epsilon_{\mathbf{k}} \hat{a}^\dagger _{\mathbf{k}, s_z} \hat{a} _{\mathbf{k}, s_z} + g \sum_{\mathbf{x}_i} \hat{n}_{\mathbf{x}_i , \uparrow} \hat{n}_{\mathbf{x}_i , \downarrow} \;.
\end{equation}
The bare interaction strength $g=V_0 / (\delta x)^2$ is fixed to reproduce the physical 2D scattering length $a$ on the lattice. The imaginary time, $\beta$, is discretized with $N_\tau$ time slices of length $\Delta\beta$ and we apply a symmetric Trotter-Suzuki decomposition,
\begin{equation}\label{trotter}
e^{-\beta \hat{H}} = \left [ e^{-\frac{\Delta\beta}{2} \hat{H}_0} e^{-\Delta\beta \hat{V}} e^{-\frac{\Delta\beta}{2} \hat{H}_0} \right ] ^{N_\tau} 
 + \mathcal{O} [ \left ( \Delta \beta \right ) ^2 ] \;,
\end{equation}
where $\hat{H}_0 = \sum_{\mathbf{k},s_z} \epsilon_{\mathbf{k}} \hat{a}^\dagger _{\mathbf{k},s_z} \hat{a}_{\mathbf{k},s_z} - g (\hat{N}_\uparrow + \hat{N}_\downarrow)/2$ is the single-particle Hamiltonian, $\hat{V} = g \sum_{\mathbf{x}} (\hat{n}_{\mathbf{x}_i,\uparrow} + \hat{n}_{\mathbf{x}_i,\downarrow}) ^2/2$ is the interaction term, and $\Delta\beta = \beta / N_\tau$. We then apply a Hubbard-Stratonovich transformation at each lattice site and time slice to decouple the interaction term,
\begin{equation}\label{HSlocal}
e^{-\Delta\beta g \hat{n}_{x_i} ^2 /2} = \sqrt{\frac{\Delta\beta |g|}{2\pi}} \int_{-\infty} ^\infty d\sigma_{\mathbf{x}_i} e^{-\Delta\beta |g| \sigma_{\mathbf{x}_i} ^2 / 2} e^{-\Delta\beta g \sigma_{\mathbf{x}_i} \hat{n}_{x_i}} \;,
\end{equation}
where $\hat{n}_{x_i} = \hat{n}_{\mathbf{x}_i,\uparrow} + \hat{n}_{\mathbf{x}_i,\downarrow}$. The introduction of the auxiliary fields $\sigma_{\mathbf{x}_i}(\tau_n)$ linearizes the quadratic interaction at the expense of a high-dimensional integral. The thermal propagator is written as a path integral over the auxiliary field configurations
\begin{equation}\label{HSThermal}
e^{-\beta \hat{H}} = \int D[\mathbf{\sigma}] G_\mathbf{\sigma} \hat{U}_\sigma \;, 
\end{equation}
where
$$
D[\mathbf{\sigma}] = \prod_{n=1} ^{N_\tau}  \prod_{\mathbf{x}_i} \left ( \sqrt{\frac{\Delta\beta |g|}{2\pi}} \right ) \textup{d}\sigma_{\mathbf{x}_i} (\tau_n)
$$
is the integration measure, $G_\mathbf{\sigma} = e^{-\Delta\beta |g| \sum_{i, n} \sigma^2 _{\mathbf{x}_i} (\tau_n) / 2}$
is a Gaussian weight, and
$$
\hat{U}_\sigma = \prod_n e^{-\Delta\beta \hat{H}_0 / 2} e^{-\Delta\beta g\sum_{\mathbf{x}_i} \mathbf{\sigma}_{\mathbf{x}_i} (\tau_n) \hat{n}_{\mathbf{x}_i}} e^{-\Delta\beta \hat{H}_0 / 2}
$$
is the thermal propagator for a given auxiliary field configuration. 

Using Eq.~(\ref{HSThermal}), the thermal expectation value of an observable $\hat O$, can then be written in the form
\begin{equation}\label{observables}
\langle \hat{O} \rangle = \frac{\Tr (\hat{O} e^{-\beta\hat{H}})}{\Tr (e^{-\beta\hat{H}})} = \frac{\int D[\mathbf{\sigma}] \langle \hat{O} \rangle_\sigma G_\mathbf{\sigma} \Tr (\hat{U}_\sigma)}{\int D[\mathbf{\sigma}] G_\mathbf{\sigma} \Tr \hat{U}_\sigma} \;.
\end{equation}
The integration over the auxiliary fields is carried out using Monte Carlo sampling. The integrals are discretized using three-point Gaussian quadrature, which has an error comparable to the Trotter-Suzuki error~\cite{Dean1993}. Since $\hat U_\sigma$ is a one-body propagator, we can evaluate the integrands using matrix algebra in the single-particle space. For example, the grand-canonical traces $\Tr \hat{U}_\sigma$ are calculated from
\begin{equation}\label{tracedeterminant}
\Tr \hat{U}_\sigma  = \det \left ( 1 + \mathbf{U}_\sigma \right),
\end{equation}
where $\mathbf{U}_\sigma$ is the matrix representation of the thermal propagator in the single-particle space. 

The canonical ensemble is implemented by projecting on a fixed particle number. We use a discrete Fourier transform~\cite{Ormand1994},
\begin{equation}\label{particlenumberprojection}
\hat{P}_{N} = \frac{e^{-\beta \mu N}}{N_s} \sum_{m = 1} ^{N_s} e^{-i \varphi_m N} e^{\left ( \beta \mu + i \varphi_m \right ) \hat{N}},
\end{equation}
where $N$ is the desired particle number, $N_s$ is the number of single-particle states, $\varphi_m = \frac{2\pi m}{N_s}$ are quadrature points, and $\mu$ is the chemical potential, introduced to ensure the numerical stability of the Fourier sum. For most calculations in this work, we perform two projections; one projection for the spin-up particles, and a second projection for the spin-down particles: $\Tr_{N_\uparrow N_\downarrow} \hat{X} = \Tr (\hat{P}_{N_{\uparrow}} \hat{P}_{N_{\downarrow}} \hat{X})$.  This canonical ensemble implementation enables the direct calculation of novel observables such as the free energy staggering pairing gap (see Sec.~\ref{sec:freegap}). We note that other projection formulas have been used in the literature~\cite{Wang2017,Shen2020}.

\begin{figure}[bth]
    \centering
    \hspace*{-1.5cm}
    \includegraphics[scale=0.3]{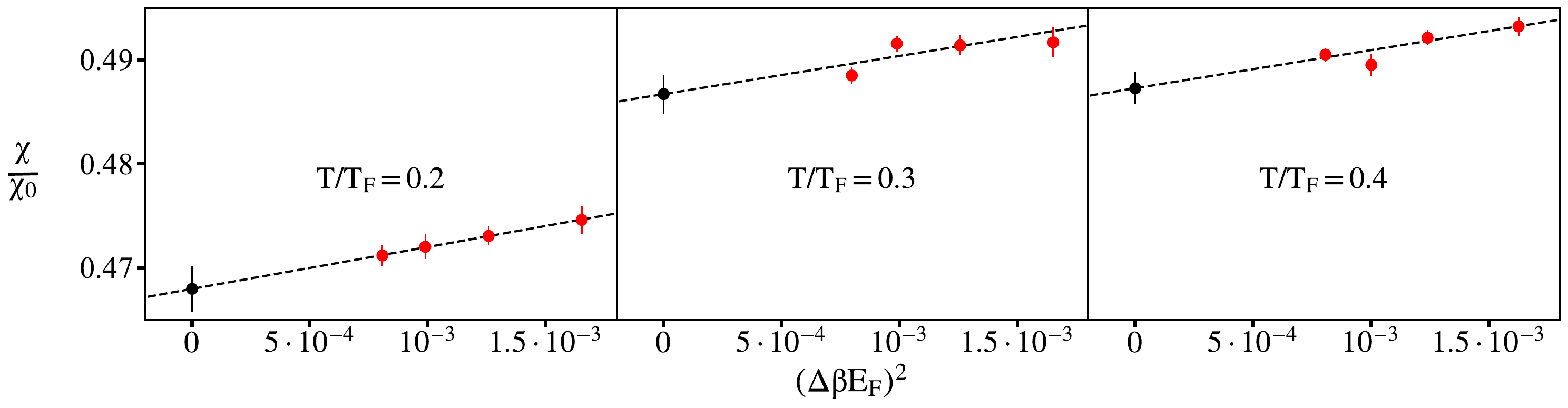}
    \caption{Continuous time extrapolations for the spin susceptibility $\chi/\chi_0$ at different temperatures for $\ln(k_F a) =1.3$ and $N=114$ particles.}
    \label{fig:chitime}
\end{figure}

The symmetric Trotter-Suzuki decomposition and three-point Gaussian quadrature introduce errors of the order of $\mathcal{O}((\Delta\beta)^2)$. To remove these errors, we take the continuous time limit $\Delta\beta\rightarrow 0$, performing a linear extrapolation in $(\Delta\beta)^2$. Typical extrapolations, for the spin susceptibility, are shown in Fig.~\ref{fig:chitime}. 

\begin{figure}[bth]
    \centering
    \hspace*{-2.0cm}
    \includegraphics[scale=0.3]{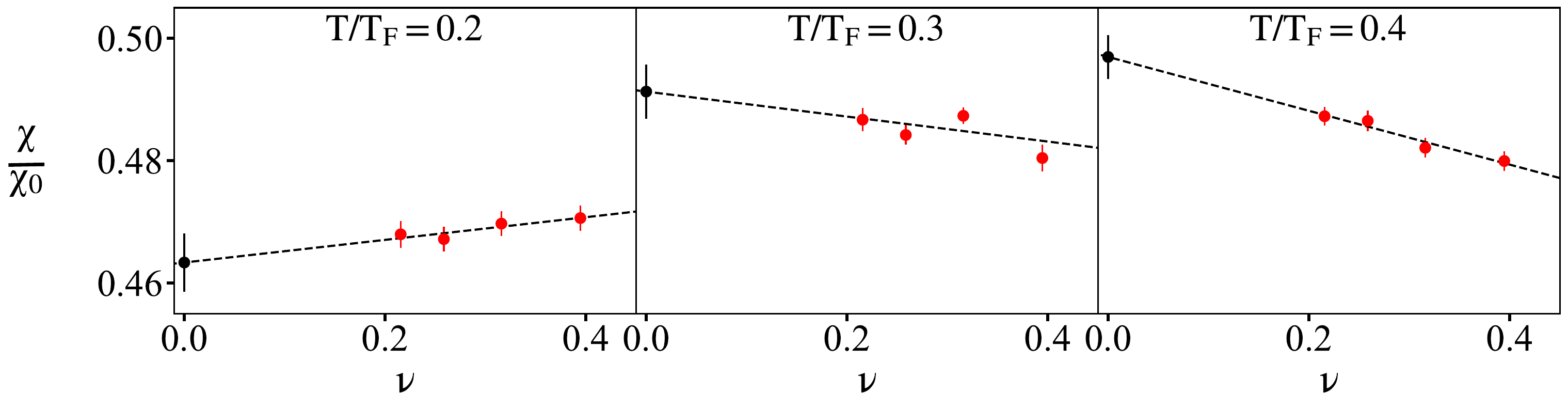}
    \caption{Continuum limit extrapolations for the spin susceptibility $\chi/\chi_0$ at different temperatures for $\ln(k_F a) =1.3$ and $N=114$ particles.}
    \label{fig:fillingtemp}
\end{figure}

The use of a finite lattice size introduces a further systematic error. To control this error, we take the continuum limit by performing a linear extrapolation in the filling factor  $\nu \to 0$~\cite{Werner2012}. An example of a typical continuum extrapolation is shown in Fig.~\ref{fig:fillingtemp} for the spin susceptibility $\chi/\chi_0$. Once the continuous time and continuum limit are reached, the only remaining errors are statistical errors for fixed particle number $N$. We then identify convergence in particle number with the thermodynamic limit.

\section{Results}

In this section we discuss results for the 2D strongly correlated Fermi gas, expanding on our recent work in Ref.~\cite{Ramachandran2024}.  In particular, we present results for the condensate fraction, spin susceptibility, energy equation of state, contact, and the free energy staggering gap.

\subsection{Condensate Fraction}
\begin{figure}
    \centering
    \hspace*{-0.5cm}
    \includegraphics[scale=0.6]{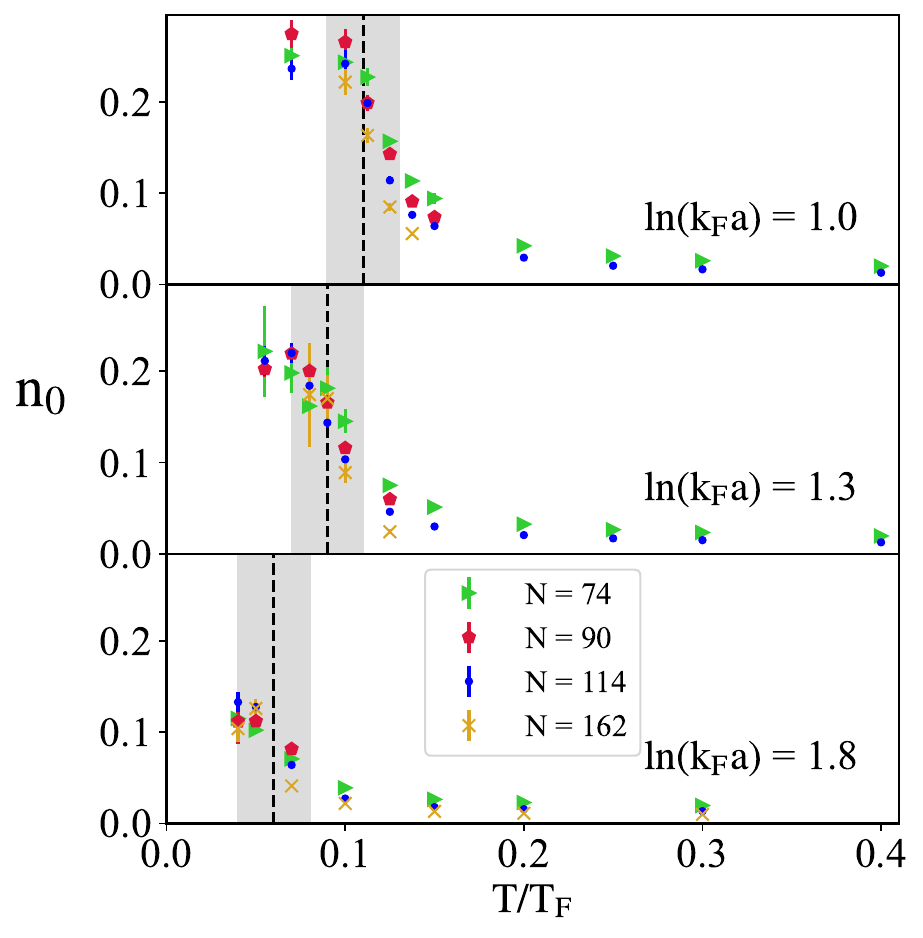}
    \caption{Condensate fraction $n_0=\lambda_{\rm max}/(N/2)$  as a function of temperature for different coupling strengths in the crossover regime. We find that above $T_c$, the condensate fraction decreases with increasing particle number, indicating the absence of a condensate in the thermodynamic limit. Below $T_c$, the condensate fraction converges to the thermodynamic limit. The condensate fraction increases as $\ln(k_F a)$ decreases towards $1.0$.}
    \label{fig:CondensateFraction}
\end{figure}

The superfluid phase transition in the 2D interacting Fermi gas belongs to the Berezinskii-Kosterlitz-Thouless (BKT) universality class. For this universality class, a condensate with off-diagonal long-range order (ODLRO)~\cite{Yang1962} exists only at zero temperature, with no ODLRO at finite temperature. Instead, below the critical temperature $T_c$, there is algebraic decay in the correlation function of the order parameter, which can be seen in the two-body density matrix~\cite{Levinsen2015} and has been observed experimentally~\cite{Ries2015}. While there is no true condensate for an infinite system, there is a macroscopic condensate fraction in a finite system. Numerically, this condensate fraction is identified with a large eigenvalue $\lambda_\mathrm{max}$ of the two-body density matrix $\langle a^\dagger _{\mathbf{k}_1 , \uparrow} a^\dagger _{\mathbf{k}_2 , \downarrow} a _{\mathbf{k}_3 , \downarrow} a _{\mathbf{k}_4 , \uparrow} \rangle$ that scales with the system size. Here, we define the condensate fraction $n_0$ by scaling $\lambda_\mathrm{max}$ with the maximal number of pairs $N/2$. In Fig.~\ref{fig:CondensateFraction}, we show the condensate fraction $n_0$ as a function of temperature for three values of the coupling strength $\ln(k_F a)$. We find that above the critical temperature $T_c$, the condensate fraction decreases with increasing particle number, indicating that this observable is sensitive to finite-size effects. In the limit $N\rightarrow\infty$, we expect that the condensate fraction would vanish above $T_c$. Below $T_c$, we find that the condensate fraction increases with decreasing coupling strength as expected.

To determine the critical temperature $T_c$, we performed in Ref.~\cite{Ramachandran2024} a finite-size scaling analysis with the condensate fraction results for multiple system sizes, for particle numbers $N= 74, 90,114$ and $162$. The finite-size scaling is based on the phenomenological renormalization group analysis of Refs.~\cite{Nightingale1982,Santos1981}.

\subsection{Spin susceptibility}
\label{sec:susceptibility}

\begin{figure*}[bth]
    \centering
    \hspace*{-0.5cm}
    \includegraphics[scale=0.25]{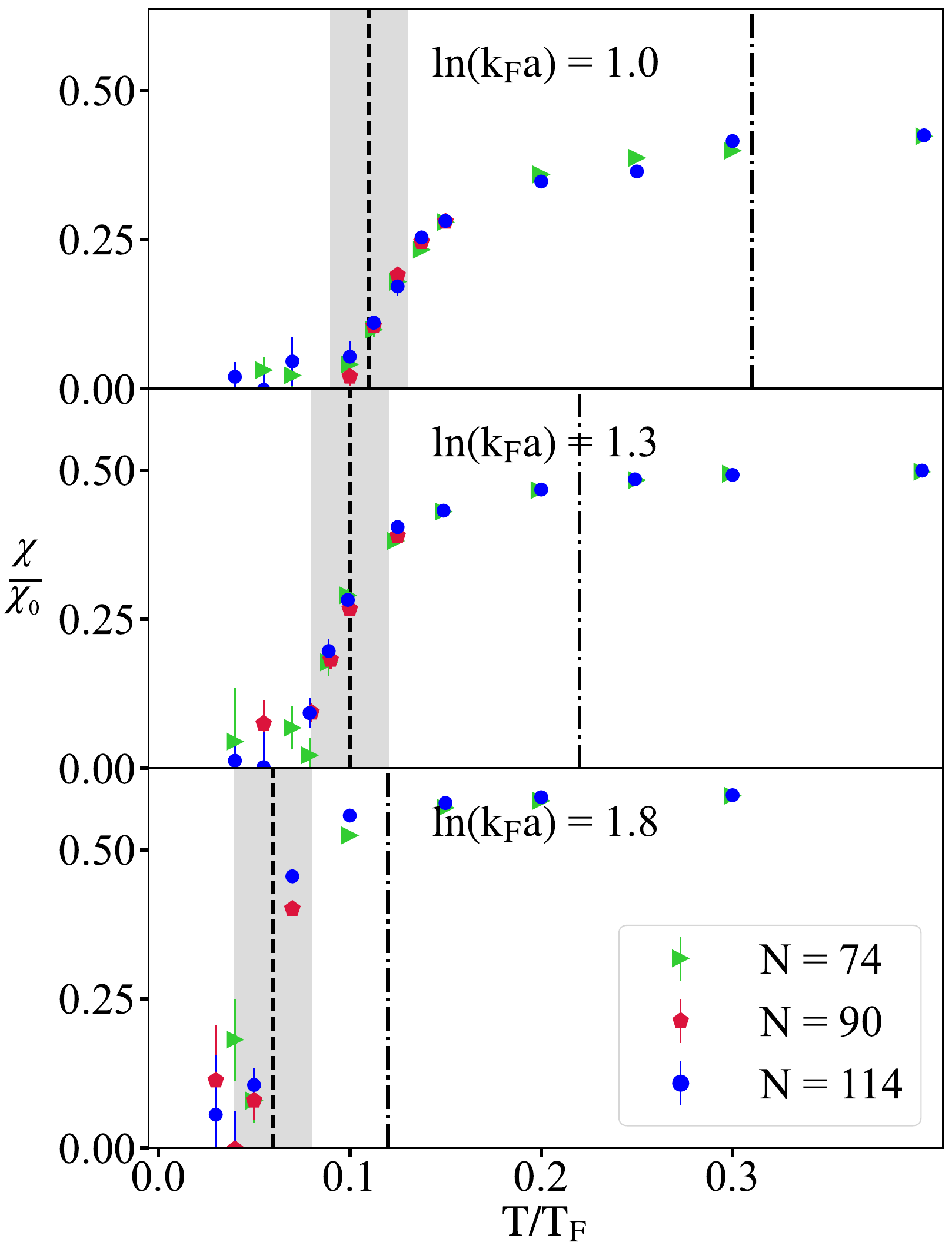}
    \caption{Spin susceptibility vs.~temperature for several coupling strengths $\ln(k_F a)$  in the crossover regime of the 2D interacting Fermi gas. We find that there is a significant pseudogap regime as $\ln(k_F a)\rightarrow 1.0$, characterized by a suppression of the spin susceptibility in a wide temperature regime. In all cases we find that the spin susceptibility is significantly suppressed in the superfluid phase below $T_c$. For each of the three interaction strengths, the black dashed line indicates the critical temperature $T_c$ determined from the FSS analysis in Ref.~\cite{Ramachandran2024}, with the associated statistical uncertainty shown as a gray band. The pseudogap temperature $T^\star$, marking the upper bound of the pseudogap regime, is shown as a dot-dashed line.}
    \label{fig:chi}
\end{figure*}

\begin{figure}[bth]
    \centering
    \hspace*{-0.5cm}
    \includegraphics[scale=0.5]{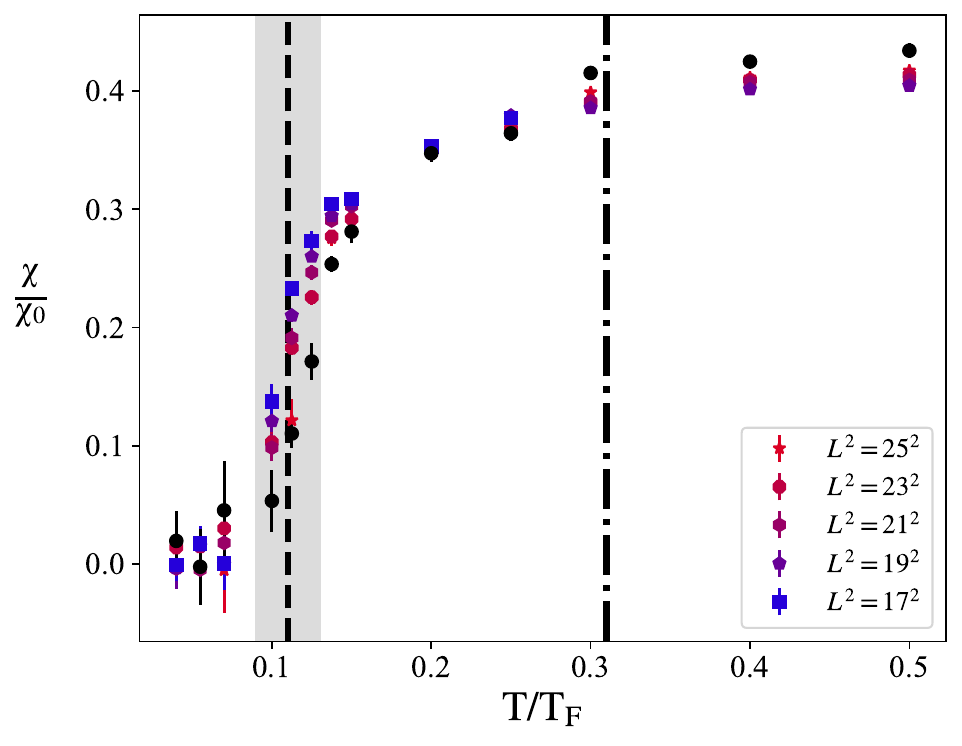}
    \caption{Spin susceptibility at $\ln(k_F a) = 1.0$ as a function of temperature for $N=114$ particles at different lattice sizes, and in the continuum limit (black circles).  }
 \label{fig:chifilling}
\end{figure}

As a probe of pairing correlations, we calculate the spin susceptibility
\[
\chi = \frac{\beta}{V} \langle (N_\uparrow - N_\downarrow) ^2 \rangle\;.
\]
For this calculation, we apply only a single particle-number projection on the total number of particles $N_\uparrow + N_\downarrow$ in the canonical ensemble AFMC framework.  For attractive Fermi gases, the $s$-wave pairing correlations suppress the spin susceptibility $\chi$. We therefore associate the suppression of $\chi$ above $T_c$ with high-temperature precursor pairing, and we identify the pseudogap regime quantitatively using this observable. We show the AFMC spin susceptibility (in units of the non-interacting spin susceptibility $\chi_0$)  as a function of temperature in Fig.~\ref{fig:chi}. We observe that below $T_c$, $\chi$ is significantly suppressed, as is expected in the superfluid phase. However, we find that there is significant suppression of $\chi$ also above $T_c$, consistent with a pronounced pseudogap regime. We define $T^\star$ to be the temperature at which $\chi$ reaches $95\%$ of its maximum value, and use it to define the upper limit of the pseudogap regime. We find that at $\ln(k_F a) = 1.0$, $T^\star / T_F = 0.31 $; at $\ln(k_F a) = 1.3$, $T^\star / T_F = 0.22$; and at $\ln(k_F a) = 1.8$, $T^\star / T_F = 0.12$. While alternative definitions of the pseudogap regime exist and may yield different values for the upper limit of the pseudogap regime, here we are providing a precise, accurate, and controlled calculation of the spin susceptibility in the continuum limit, approaching the thermodynamic limit for this strongly correlated system. This robust result offers a clear indirect evidence of pairing correlations for temperatures above $T_c$.

In Fig.~\ref{fig:chifilling}, we show the spin susceptibility as a function of temperature for several lattice sizes, along with the continuum results (black circles). We find that for $T>T^\star$, the finite lattice calculations underestimate the continuum limit values of $\chi$, while in the pseudogap regime $T_c<T<T^\star$, the finite lattice calculations overestimate $\chi$. As a result, we find that the pseudogap effects are enhanced in the continuum limit, making the continuum extrapolation necessary for an accurate quantitative definition of the pseudogap regime. 

\subsection{Energy equation of state}

\begin{figure*}[bth]
    \centering
    \hspace*{-0.5cm}
    \includegraphics[scale=0.5]{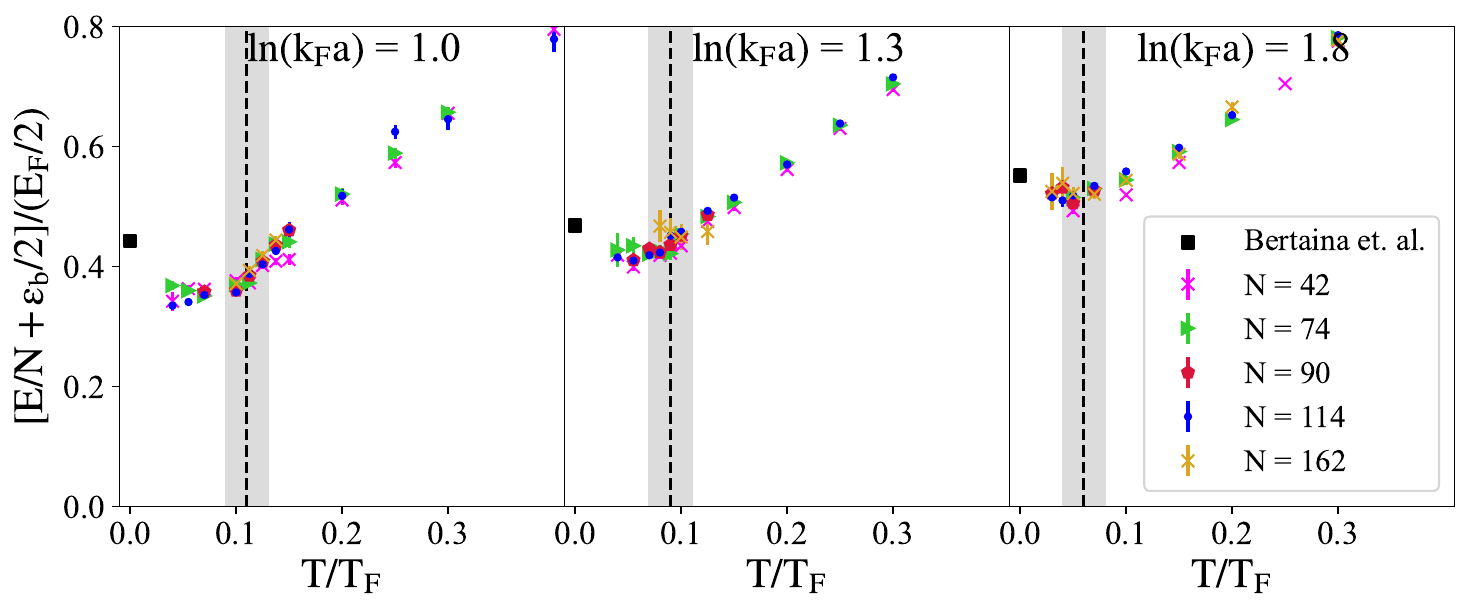}
    \caption{Energy equation of state vs.~temperature for different coupling strengths. We compare with the $T=0$ diffusion Monte Carlo results of Ref.~\cite{Bertaina2011} (black squares).  The results at $\ln(k_F a) = 1.3$ and $\ln(k_F a) = 1.8$ are obtained from the fit in Ref.~\cite{Bertaina2011}.  We also find that the energy at low temperatures is significantly lower than the $T=0$ result of Ref.~\cite{Bertaina2011} at $\ln(k_F a) = 1.0$.}
    \label{fig:EEOS}
\end{figure*}

We calculate the thermal energy $E$  as the expectation value of the Hamiltonian with respect to the Gibbs density matrix at finite temperature. As in Ref.~\cite{Bertaina2011}, we define the energy equation of state to be $E/N + \epsilon_b/2$ (in units of $E_F/2$) vs.~$T/T_F$, where $\epsilon_b$ is the two-particle bound state energy. Our AFMC results are shown in Fig.~\ref{fig:EEOS} and compared with the $T=0$ diffusion Monte Carlo results of Ref.~\cite{Bertaina2011} (black squares).

We find that this energy depends weakly on temperature in the superfluid phase but increases linearly with temperature above $T_c$. At $\ln(k_F a) = 1.8$, the energy at low temperatures is in overall agreement with $T=0$ diffusion Monte Carlo results of Ref.~\cite{Bertaina2011}.  As the coupling strength decreases, the energy of the BKT phase decreases, and at $\ln(k_F a) = 1.0$,  the AFMC energy at low temperatures is significantly smaller than the energy found in Ref.~\cite{Bertaina2011}.

\subsection{Contact}

\begin{figure}[bth]
    \centering
    \hspace*{-0.5cm}
    \includegraphics[scale=0.5]{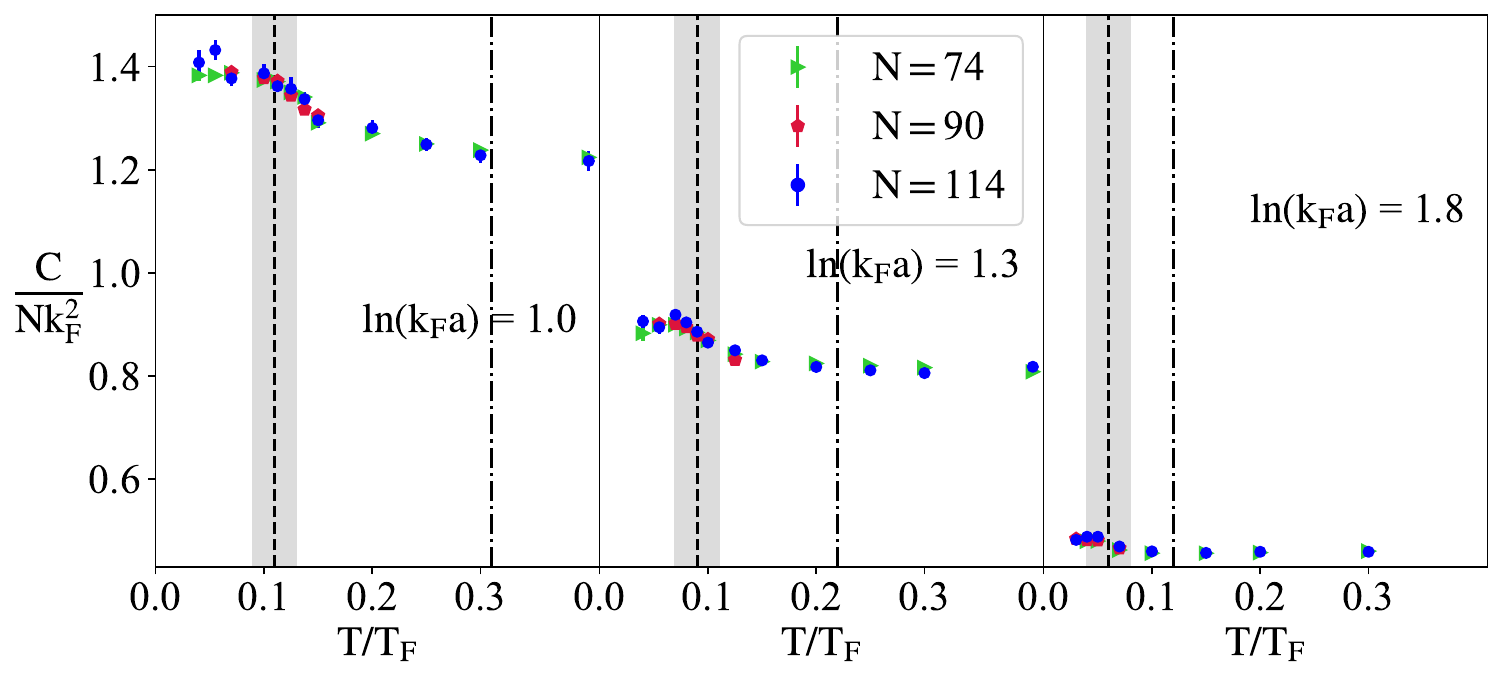}
    \caption{Contact vs.~temperature for different coupling strengths. We find that the contact increases with decreasing temperature in the pseudogap regime.
     We also find that the contact increases with decreasing $\ln(k_F a)$.}
    \label{fig:ContactvT}
\end{figure}

The contact $C$ is a measure of short-range correlations between particles of opposite spin. It is defined by
\begin{equation}
\int d^2 R \ g^{(2)} _{\uparrow\downarrow} \left ( \mathbf{R} + \frac{\mathbf{r}}{2} , \mathbf{R} - \frac{\mathbf{r}}{2} \right ) \underset{r\to 0}{\sim} \frac{C}{(2\pi)^2} \ln ^2 (r) \;,
\end{equation}
where $g^{(2)} _{\uparrow\downarrow} \left ( \mathbf{R} + \frac{\mathbf{r}}{2} , \mathbf{R} - \frac{\mathbf{r}}{2} \right ) = \langle \hat{n}_\uparrow \left ( \mathbf{r}_\uparrow \right ) \hat{n}_\downarrow \left ( \mathbf{r}_\downarrow \right ) \rangle$ is the two-body correlation function. The contact has emerged as a key observable of interest due to Tan's relations, which relate the contact to universal properties of interacting Fermi gases~\cite{Tan2008,Werner2012}. For instance, the contact describes the behavior of the momentum distribution at large $k$, $n_\sigma (\mathbf{k})\sim C/k^4$. The contact can also be expressed as the derivative of the energy with respect to the coupling strength
\begin{equation}
\frac{\hbar^2 C}{2\pi m} = \frac{d E}{d \ln a} \;.
\end{equation}
We use this relation in our lattice calculations to calculate $C$ from the thermal potential energy~\cite{Jensen2020Contact}
\begin{equation}
C = \frac{m^2}{\hbar^4} V_0 \langle \hat{V} \rangle \;.
\end{equation}
In Fig.~\ref{fig:ContactvT} we show our results for the contact (in units of $N k_F^2$) as a function of temperature (in units of $T_F$). For all three coupling strengths, the contact increases as $T$ decreases below $T_c$. We also find that the contact monotonically increases with decreasing temperature in the pseudogap regime, $T_c < T < T^\star$. This increase is largest at $\ln(k_F a) \sim 1.0$, for which we observe the largest pseudogap signatures.

\begin{figure}[h!]
    \centering
    \hspace*{-0.5cm}
    \includegraphics[scale=0.5]{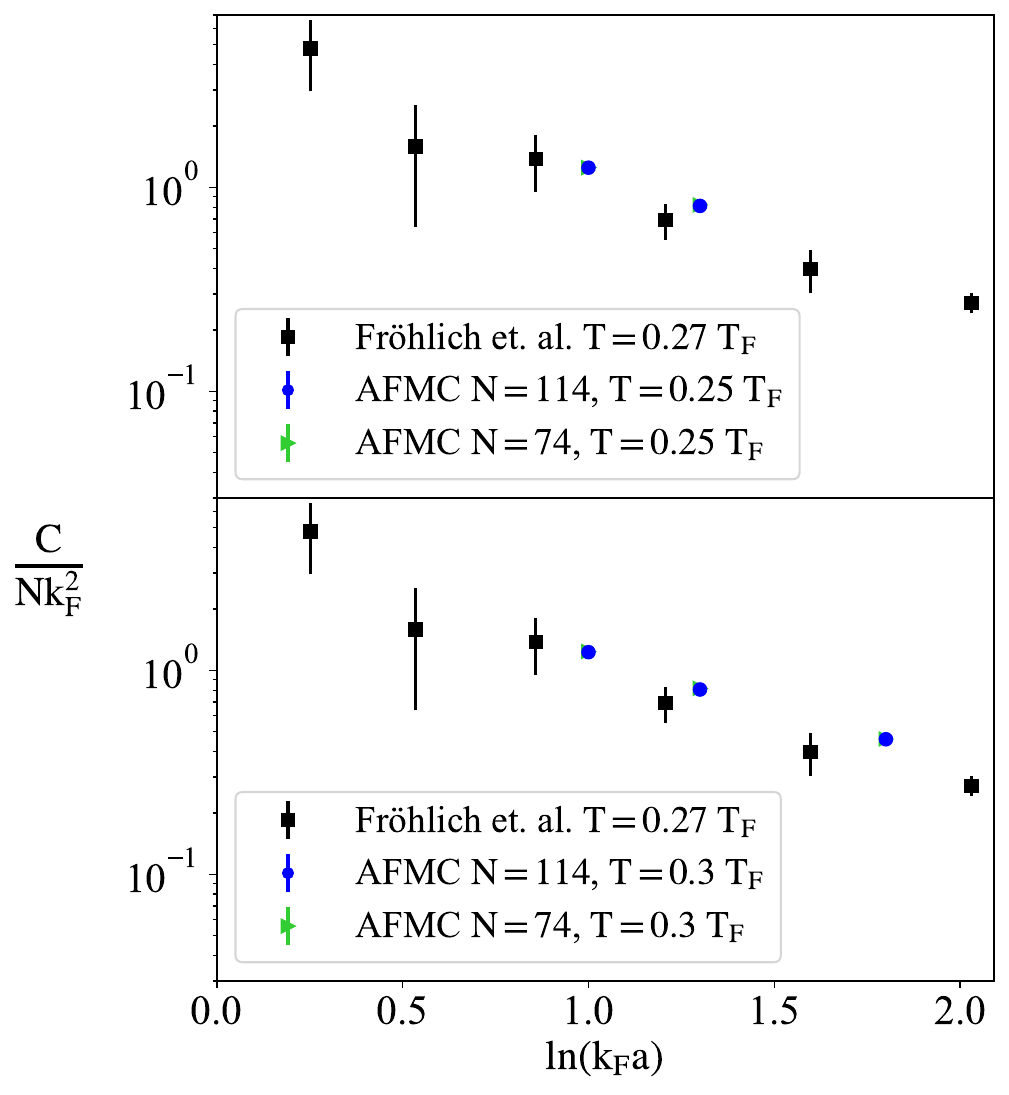}
    \caption{Contact vs.~coupling strength in the neighborhood of $T / T_F \sim 0.27$. We compare our AFMC results at $T = 0.25 \,T_F$ (top panel) and $T = 0.3 \,T_F$ (bottom panel) with the experimental results of Ref.~\cite{Frolich2012} at $T=0.27\, T_F$.}
    \label{fig:ContactvEta}
\end{figure}

We also compare in Fig.~\ref{fig:ContactvEta} our results for the contact as a function of the coupling strength $\ln(k_F a)$ to select experimental results of Ref.~\cite{Frolich2012} at $T/T_F = 0.27$. 

\subsection{Free energy gap}
\label{sec:freegap}

\begin{figure}[bth]
    \centering
    \hspace*{-0.5cm}
    \includegraphics[scale=0.35]{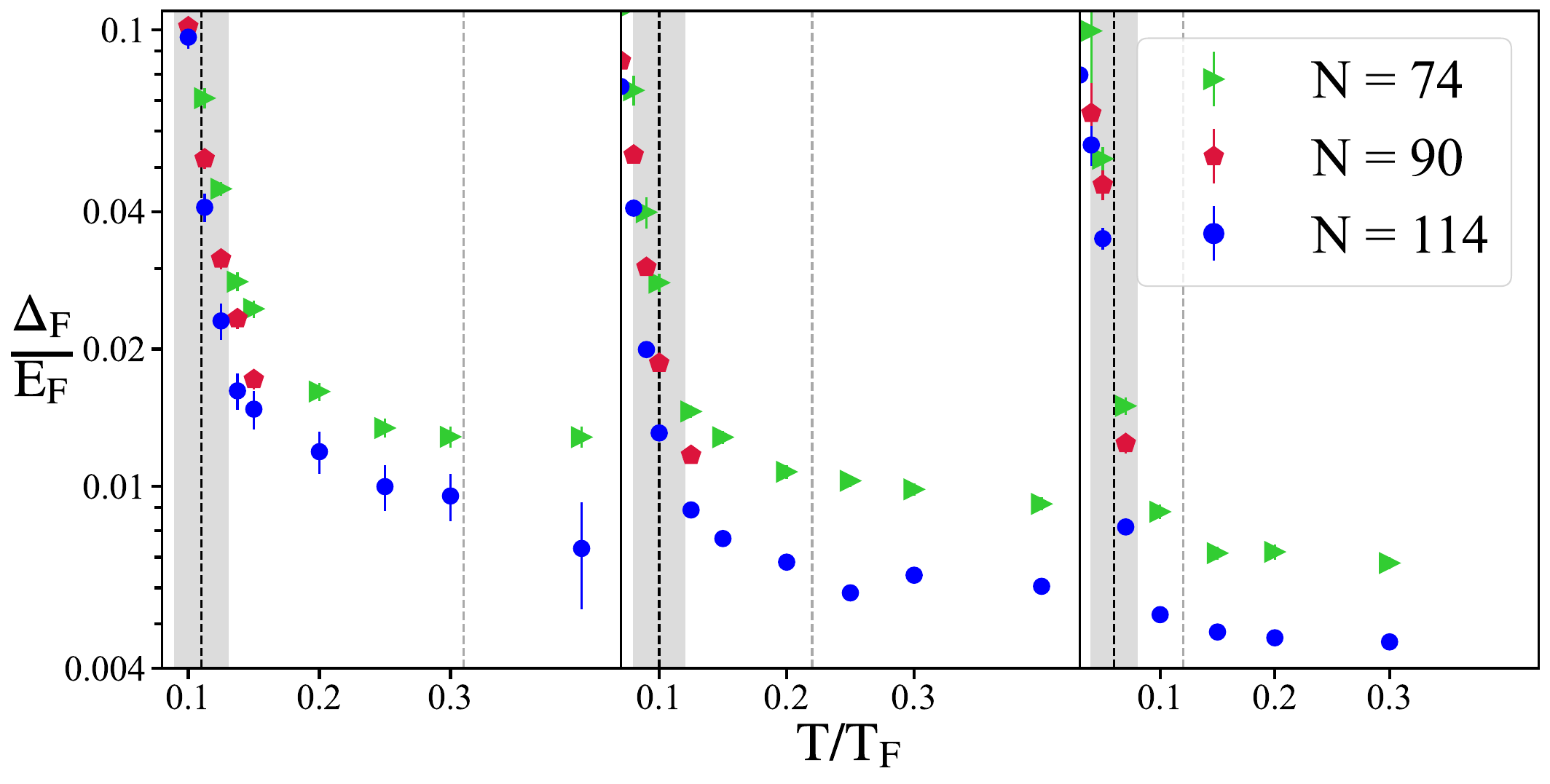}
    \caption{Free energy gap $\Delta_F$ as a function of temperature $T/T_F$, plotted on a logarithmic scale. We find that  $\Delta_F$ increases with decreasing temperature in the pseudogap regime.}
    \label{fig:FGapvsT}
\end{figure}

Calculating the energy staggering pairing gap with AFMC in two dimensions is difficult due to a sign problem that emerges when considering deviations from spin balance. Instead, we calculate the free energy gap, defined as
\begin{equation}
\Delta_F = [2F\left(N/2-1,N/2\right)-F\left (N/2-1,N/2-1\right)\\-F\left(N/2,N/2\right)] \;.
\end{equation}
We show calculations for the free energy gap above $T_c$ in Fig.~\ref{fig:FGapvsT}. When plotted on a logarithmic scale, we find that the free energy gap increases with decreasing temperature in the spin gap regime, serving as a pseudogap signature. In this regime, we also find that the free energy gap does not converge in particle number, instead displaying a systematic decrease with increasing particle number, similar to the condensate fraction. In Ref.~~\cite{Ramachandran2024} we found that at low temperatures, the free energy gap displays a linear dependence on temperature and used a linear extrapolation to the $T=0$ limit, where the free energy gap is equal to the energy staggering pairing gap. This provides us with an accurate way to determine the $T=0$ energy staggering pairing gap.

\section{Conclusions and Outlook}

In this work we have investigated thermal properties of the two-dimensional BCS-BEC crossover in the strongly correlated regime, $\ln(k_Fa) \sim1$, using continuum limit AFMC calculations. We presented results for the condensate fraction, building on the work in Ref.~\cite{Ramachandran2024}. We also calculated the energy of the system and Tan's contact, obtaining accurate and controlled results for the thermodynamic behavior of the system. We note that our results for the energy equation of state at low temperatures are considerably lower than those of Ref.~\cite{Bertaina2011}.

Through calculations of the spin susceptibility and the free energy gap with canonical ensemble AFMC in the continuum limit, we identified the emergence of a pseudogap regime with controlled calculations as $\ln(k_F a)\rightarrow 1$ from above. In particular, we find that the spin susceptibility $\chi$ is suppressed for a large range of temperatures above $T_c$. We also found that the contact and the free energy gap increase monotonically with decreasing temperature in the strongly correlated regime.

Our results indicate of a pronounced pseudogap regime along the two-dimensional BCS-BEC crossover. However, due to the presence of a bound state for an arbitrarily weak attractive interaction, it is unclear how much of the pseudogap phenomenon can be attributed to two-particle pairing and how much is due to many-body precursor pairing correlations. To address many-body pairing correlations, we focused our calculations on the BCS side where the two-particle bound state does not dominate, i.e., for $\ln(k_F a)\rightarrow 1$ from above. Our estimates for the pairing temperature scale $T^\star$ are $T^\star = 0.91 |\epsilon_b|$ at $\ln(k_F a) = 1.0$, $T^\star = 1.17 |\epsilon_b|$ at $\ln(k_F a) = 1.3$,  and $T^\star = 1.74 |\epsilon_b|$ at $\ln(k_F a) = 1.8$, where $\epsilon_b$ is the two-particle binding energy. These estimates suggest that the two-particle binding energy becomes increasingly relevant towards the BEC limit, where the pseudogap scale at $\ln(k_F a) = 1$ is close to the scale of the binding energy. 

A recent experimental study found high-temperature phenomena attributed to many-body pairing at $\ln k_F a \sim 1$ and $T/T_F \sim 0.5$~\cite{Murthy2018}, where a pairing gap was observed to significantly exceed the two-particle binding energy. 
Here, for the same interaction and temperature regime, we estimate $T^\star = 0.31\, T_F$ (using the criterion discussed in Sec.~\ref{sec:susceptibility}), with no significant pseudogap effects at $T/T_F \sim 0.5$ for $\ln k_F a \sim 1$. A future study of the spectral function with controlled AFMC calculations will allow for a direct comparison with Ref.~\cite{Murthy2018} and provide valuable benchmark results for future experimental investigations.

Interest in the two-dimensional BCS-BEC crossover has been steadily growing with new experimental breakthroughs allowing for precision many-body studies of this strongly correlated system. Near future work will likely focus on elucidating the equation of state. The two-dimensional equation of state at spin balance has been explored both theoretically~\cite{Bauer2014,Anderson2015} and experimentally~\cite{Boettcher2016,Fenech2016}, with a considerable quantitative discrepancy.  The phase diagram of the  spin-imbalanced system is still an open problem. Of particular interest is the possible presence of a Fulde-Ferrell-Larkin-Ovchinnikov (FFLO) phase~\cite{Fulde1964,Larkin1965}, especially in the presence of spin-orbit coupling~\cite{Shi2016, Vitali2022, Rosenberg2017, Chen2012, Zheng2014}. 

\section{Acknowledgements}  This work was supported in part by the U.S. DOE grants No.~DE-SC0019521 and No.~DE-SC0020177. The calculations used resources of the National Energy Research Scientific Computing Center (NERSC), a U.S. Department of Energy Office of Science User Facility operated under Contract No.~DE-AC02-05CH11231.  We thank the Yale Center for Research Computing for guidance and use of the research computing infrastructure.

The data files and scripts used to generate the figures in this work are provided in the Electronic Supplemental Material.


\begin{thebibliography}{10}
\expandafter\ifx\csname url\endcsname\relax
  \def\url#1{\burl{#1}}\fi
\expandafter\ifx\csname urlprefix\endcsname\relax\def\urlprefix{URL }\fi
\providecommand{\bibinfo}[2]{#2}
\providecommand{\eprint}[2][]{\url{#2}}
\providecommand{\doi}[1]{\url{https://doi.org/#1}}
\bibcommenthead

\bibitem{Jensen2020Pairing}
\bibinfo{author}{Jensen, S.}, \bibinfo{author}{Gilbreth, C.~N.} \&
  \bibinfo{author}{Alhassid, Y.}
\newblock \bibinfo{title}{Pairing correlations across the superfluid phase
  transition in the unitary {Fermi }gas}.
\newblock \emph{\bibinfo{journal}{Phys. Rev. Lett.}}
  \textbf{\bibinfo{volume}{124}}, \bibinfo{pages}{090604}
  (\bibinfo{year}{2020}).
\newblock
  \urlprefix\url{https://link.aps.org/doi/10.1103/PhysRevLett.124.090604}.

\bibitem{Jensen2024}
\bibinfo{author}{Jensen, S.}, \bibinfo{author}{Gilbreth, C.~N.} \&
  \bibinfo{author}{Alhassid, Y.}
\newblock \bibinfo{title}{Pseudogap regime of the unitary {Fermi }gas with
  lattice auxiliary-field quantum {Monte Carlo} in the continuum limit}
  (\bibinfo{year}{2024}).
\newblock \urlprefix\url{https://arxiv.org/abs/2408.16676}.
\newblock
  \bibinfo{eprint}{{\href{https://arxiv.org/abs/2408.16676}{{arXiv:2408.16676}}}}.

\bibitem{Frolich2011}
\bibinfo{author}{Fr\"ohlich, B.} \emph{et~al.}
\newblock \bibinfo{title}{Radio-frequency spectroscopy of a strongly
  interacting two-dimensional {Fermi }gas}.
\newblock \emph{\bibinfo{journal}{Phys. Rev. Lett.}}
  \textbf{\bibinfo{volume}{106}}, \bibinfo{pages}{105301}
  (\bibinfo{year}{2011}).
\newblock
  \urlprefix\url{https://link.aps.org/doi/10.1103/PhysRevLett.106.105301}.

\bibitem{Vogt2012}
\bibinfo{author}{Vogt, E.} \emph{et~al.}
\newblock \bibinfo{title}{Scale invariance and viscosity of a two-dimensional
  {Fermi }gas}.
\newblock \emph{\bibinfo{journal}{Phys. Rev. Lett.}}
  \textbf{\bibinfo{volume}{108}}, \bibinfo{pages}{070404}
  (\bibinfo{year}{2012}).
\newblock
  \urlprefix\url{https://link.aps.org/doi/10.1103/PhysRevLett.108.070404}.

\bibitem{Murthy2015}
\bibinfo{author}{Murthy, P.~A.} \emph{et~al.}
\newblock \bibinfo{title}{Observation of the
  {Berezinskii}-{Kosterlitz}-{Thouless} phase transition in an ultracold {Fermi
  }gas}.
\newblock \emph{\bibinfo{journal}{Phys. Rev. Lett.}}
  \textbf{\bibinfo{volume}{115}}, \bibinfo{pages}{010401}
  (\bibinfo{year}{2015}).
\newblock
  \urlprefix\url{https://link.aps.org/doi/10.1103/PhysRevLett.115.010401}.

\bibitem{Ries2015}
\bibinfo{author}{Ries, M.~G.} \emph{et~al.}
\newblock \bibinfo{title}{Observation of pair condensation in the quasi-{2D}
  {BEC-BCS} crossover}.
\newblock \emph{\bibinfo{journal}{Phys. Rev. Lett.}}
  \textbf{\bibinfo{volume}{114}}, \bibinfo{pages}{230401}
  (\bibinfo{year}{2015}).
\newblock
  \urlprefix\url{https://link.aps.org/doi/10.1103/PhysRevLett.114.230401}.

\bibitem{Boettcher2016}
\bibinfo{author}{Boettcher, I.} \emph{et~al.}
\newblock \bibinfo{title}{Equation of state of ultracold fermions in the {2D}
  {BEC-BCS} crossover region}.
\newblock \emph{\bibinfo{journal}{Phys. Rev. Lett.}}
  \textbf{\bibinfo{volume}{116}}, \bibinfo{pages}{045303}
  (\bibinfo{year}{2016}).
\newblock
  \urlprefix\url{https://link.aps.org/doi/10.1103/PhysRevLett.116.045303}.

\bibitem{Toniolo2017}
\bibinfo{author}{Toniolo, U.}, \bibinfo{author}{Mulkerin, B.~C.},
  \bibinfo{author}{Vale, C.~J.}, \bibinfo{author}{Liu, X.-J.} \&
  \bibinfo{author}{Hu, H.}
\newblock \bibinfo{title}{Dimensional crossover in a strongly interacting
  ultracold atomic {Fermi }gas}.
\newblock \emph{\bibinfo{journal}{Phys. Rev. A}} \textbf{\bibinfo{volume}{96}},
  \bibinfo{pages}{041604} (\bibinfo{year}{2017}).
\newblock \urlprefix\url{https://link.aps.org/doi/10.1103/PhysRevA.96.041604}.

\bibitem{Luciuk2017}
\bibinfo{author}{Luciuk, C.} \emph{et~al.}
\newblock \bibinfo{title}{Observation of quantum-limited spin transport in
  strongly interacting two-dimensional {Fermi }gases}.
\newblock \emph{\bibinfo{journal}{Phys. Rev. Lett.}}
  \textbf{\bibinfo{volume}{118}}, \bibinfo{pages}{130405}
  (\bibinfo{year}{2017}).
\newblock
  \urlprefix\url{https://link.aps.org/doi/10.1103/PhysRevLett.118.130405}.

\bibitem{Hueck2018}
\bibinfo{author}{Hueck, K.} \emph{et~al.}
\newblock \bibinfo{title}{Two-dimensional homogeneous {Fermi }gases}.
\newblock \emph{\bibinfo{journal}{Phys. Rev. Lett.}}
  \textbf{\bibinfo{volume}{120}}, \bibinfo{pages}{060402}
  (\bibinfo{year}{2018}).
\newblock
  \urlprefix\url{https://link.aps.org/doi/10.1103/PhysRevLett.120.060402}.

\bibitem{Murthy2018}
\bibinfo{author}{Murthy, P.~A.} \emph{et~al.}
\newblock \bibinfo{title}{High-temperature pairing in a strongly interacting
  two-dimensional {Fermi }gas}.
\newblock \emph{\bibinfo{journal}{Science}} \textbf{\bibinfo{volume}{359}},
  \bibinfo{pages}{452--455} (\bibinfo{year}{2018}).
\newblock
  \urlprefix\url{https://www.science.org/doi/abs/10.1126/science.aan5950}.

\bibitem{Feld2011}
\bibinfo{author}{Feld, M.}, \bibinfo{author}{Fr{\"o}hlich, B.},
  \bibinfo{author}{Vogt, E.}, \bibinfo{author}{Koschorreck, M.} \&
  \bibinfo{author}{K{\"o}hl, M.}
\newblock \bibinfo{title}{Observation of a pairing pseudogap in a
  two-dimensional {Fermi }gas}.
\newblock \emph{\bibinfo{journal}{Nature}} \textbf{\bibinfo{volume}{480}},
  \bibinfo{pages}{75--78} (\bibinfo{year}{2011}).
\newblock \urlprefix\url{https://doi.org/10.1038/nature10627}.

\bibitem{Watanabe2013}
\bibinfo{author}{Watanabe, R.}, \bibinfo{author}{Tsuchiya, S.} \&
  \bibinfo{author}{Ohashi, Y.}
\newblock \bibinfo{title}{Two-dimensional pseudogap effects of an ultracold
  {Fermi }gas in the {BCS-BEC} crossover region}.
\newblock \emph{\bibinfo{journal}{Journal of Low Temperature Physics}}
  \textbf{\bibinfo{volume}{171}}, \bibinfo{pages}{341--347}
  (\bibinfo{year}{2013}).
\newblock \urlprefix\url{https://doi.org/10.1007/s10909-012-0691-7}.

\bibitem{Matsumoto2014}
\bibinfo{author}{Matsumoto, M.} \& \bibinfo{author}{Ohashi, Y.}
\newblock \bibinfo{title}{Pseudogap phenomena in a two-dimensional ultracold
  {Fermi }gas near the {Berezinskii}-{Kosterlitz}-{Thouless} transition}.
\newblock \emph{\bibinfo{journal}{Journal of Physics: Conference Series}}
  \textbf{\bibinfo{volume}{568}}, \bibinfo{pages}{012012}
  (\bibinfo{year}{2014}).
\newblock \urlprefix\url{https://dx.doi.org/10.1088/1742-6596/568/1/012012}.

\bibitem{Bauer2014}
\bibinfo{author}{Bauer, M.}, \bibinfo{author}{Parish, M.~M.} \&
  \bibinfo{author}{Enss, T.}
\newblock \bibinfo{title}{Universal equation of state and pseudogap in the
  two-dimensional {Fermi }gas}.
\newblock \emph{\bibinfo{journal}{Phys. Rev. Lett.}}
  \textbf{\bibinfo{volume}{112}}, \bibinfo{pages}{135302}
  (\bibinfo{year}{2014}).
\newblock
  \urlprefix\url{https://link.aps.org/doi/10.1103/PhysRevLett.112.135302}.

\bibitem{Anderson2015}
\bibinfo{author}{Anderson, E.~R.} \& \bibinfo{author}{Drut, J.~E.}
\newblock \bibinfo{title}{Pressure, compressibility, and contact of the
  two-dimensional attractive {Fermi }gas}.
\newblock \emph{\bibinfo{journal}{Phys. Rev. Lett.}}
  \textbf{\bibinfo{volume}{115}}, \bibinfo{pages}{115301}
  (\bibinfo{year}{2015}).
\newblock
  \urlprefix\url{https://link.aps.org/doi/10.1103/PhysRevLett.115.115301}.

\bibitem{Marsiglio2015}
\bibinfo{author}{Marsiglio, F.}, \bibinfo{author}{Pieri, P.},
  \bibinfo{author}{Perali, A.}, \bibinfo{author}{Palestini, F.} \&
  \bibinfo{author}{Strinati, G.~C.}
\newblock \bibinfo{title}{Pairing effects in the normal phase of a
  two-dimensional {Fermi }gas}.
\newblock \emph{\bibinfo{journal}{Phys. Rev. B}} \textbf{\bibinfo{volume}{91}},
  \bibinfo{pages}{054509} (\bibinfo{year}{2015}).
\newblock \urlprefix\url{https://link.aps.org/doi/10.1103/PhysRevB.91.054509}.

\bibitem{Shi2015}
\bibinfo{author}{Shi, H.}, \bibinfo{author}{Chiesa, S.} \&
  \bibinfo{author}{Zhang, S.}
\newblock \bibinfo{title}{Ground-state properties of strongly interacting
  {Fermi }gases in two dimensions}.
\newblock \emph{\bibinfo{journal}{Phys. Rev. A}} \textbf{\bibinfo{volume}{92}},
  \bibinfo{pages}{033603} (\bibinfo{year}{2015}).
\newblock \urlprefix\url{https://link.aps.org/doi/10.1103/PhysRevA.92.033603}.

\bibitem{Galea2016}
\bibinfo{author}{Galea, A.}, \bibinfo{author}{Dawkins, H.},
  \bibinfo{author}{Gandolfi, S.} \& \bibinfo{author}{Gezerlis, A.}
\newblock \bibinfo{title}{Diffusion {Monte Carlo} study of strongly interacting
  two-dimensional {Fermi }gases}.
\newblock \emph{\bibinfo{journal}{Phys. Rev. A}} \textbf{\bibinfo{volume}{93}},
  \bibinfo{pages}{023602} (\bibinfo{year}{2016}).
\newblock \urlprefix\url{https://link.aps.org/doi/10.1103/PhysRevA.93.023602}.

\bibitem{Vitali2017}
\bibinfo{author}{Vitali, E.}, \bibinfo{author}{Shi, H.}, \bibinfo{author}{Qin,
  M.} \& \bibinfo{author}{Zhang, S.}
\newblock \bibinfo{title}{Visualizing the {BEC-BCS} crossover in a
  two-dimensional {Fermi }gas: pairing gaps and dynamical response functions
  from ab initio computations}.
\newblock \emph{\bibinfo{journal}{Phys. Rev. A}} \textbf{\bibinfo{volume}{96}},
  \bibinfo{pages}{061601} (\bibinfo{year}{2017}).
\newblock \urlprefix\url{https://link.aps.org/doi/10.1103/PhysRevA.96.061601}.

\bibitem{Madeira2017}
\bibinfo{author}{Madeira, L.}, \bibinfo{author}{Gandolfi, S.} \&
  \bibinfo{author}{Schmidt, K.~E.}
\newblock \bibinfo{title}{Core structure of two-dimensional {Fermi }gas
  vortices in the {BEC-BCS} crossover region}.
\newblock \emph{\bibinfo{journal}{Phys. Rev. A}} \textbf{\bibinfo{volume}{95}},
  \bibinfo{pages}{053603} (\bibinfo{year}{2017}).
\newblock \urlprefix\url{https://link.aps.org/doi/10.1103/PhysRevA.95.053603}.

\bibitem{Schonenberg2017}
\bibinfo{author}{Schonenberg, L.~M.}, \bibinfo{author}{Verpoort, P.~C.} \&
  \bibinfo{author}{Conduit, G.~J.}
\newblock \bibinfo{title}{Effective-range dependence of two-dimensional {Fermi
  }gases}.
\newblock \emph{\bibinfo{journal}{Phys. Rev. A}} \textbf{\bibinfo{volume}{96}},
  \bibinfo{pages}{023619} (\bibinfo{year}{2017}).
\newblock \urlprefix\url{https://link.aps.org/doi/10.1103/PhysRevA.96.023619}.

\bibitem{Mulkerin2018}
\bibinfo{author}{Mulkerin, B.~C.}, \bibinfo{author}{Liu, X.-J.} \&
  \bibinfo{author}{Hu, H.}
\newblock \bibinfo{title}{Collective modes of a two-dimensional {Fermi }gas at
  finite temperature}.
\newblock \emph{\bibinfo{journal}{Phys. Rev. A}} \textbf{\bibinfo{volume}{97}},
  \bibinfo{pages}{053612} (\bibinfo{year}{2018}).
\newblock \urlprefix\url{https://link.aps.org/doi/10.1103/PhysRevA.97.053612}.

\bibitem{Hu2019}
\bibinfo{author}{Hu, H.}, \bibinfo{author}{Mulkerin, B.~C.},
  \bibinfo{author}{Toniolo, U.}, \bibinfo{author}{He, L.} \&
  \bibinfo{author}{Liu, X.-J.}
\newblock \bibinfo{title}{Reduced quantum anomaly in a quasi-two-dimensional
  {Fermi }superfluid: significance of the confinement-induced effective range
  of interactions}.
\newblock \emph{\bibinfo{journal}{Phys. Rev. Lett.}}
  \textbf{\bibinfo{volume}{122}}, \bibinfo{pages}{070401}
  (\bibinfo{year}{2019}).
\newblock
  \urlprefix\url{https://link.aps.org/doi/10.1103/PhysRevLett.122.070401}.

\bibitem{Wu2020}
\bibinfo{author}{Wu, F.}, \bibinfo{author}{Hu, J.}, \bibinfo{author}{He, L.},
  \bibinfo{author}{Liu, X.-J.} \& \bibinfo{author}{Hu, H.}
\newblock \bibinfo{title}{Effective theory for ultracold strongly interacting
  fermionic atoms in two dimensions}.
\newblock \emph{\bibinfo{journal}{Phys. Rev. A}}
  \textbf{\bibinfo{volume}{101}}, \bibinfo{pages}{043607}
  (\bibinfo{year}{2020}).
\newblock \urlprefix\url{https://link.aps.org/doi/10.1103/PhysRevA.101.043607}.

\bibitem{Pascucci2020}
\bibinfo{author}{Pascucci, F.} \& \bibinfo{author}{Salasnich, L.}
\newblock \bibinfo{title}{Josephson effect with superfluid fermions in the
  two-dimensional {BCS-BEC} crossover}.
\newblock \emph{\bibinfo{journal}{Phys. Rev. A}}
  \textbf{\bibinfo{volume}{102}}, \bibinfo{pages}{013325}
  (\bibinfo{year}{2020}).
\newblock \urlprefix\url{https://link.aps.org/doi/10.1103/PhysRevA.102.013325}.

\bibitem{Zhao2020}
\bibinfo{author}{Zhao, H.}, \bibinfo{author}{Gao, X.}, \bibinfo{author}{Liang,
  W.}, \bibinfo{author}{Zou, P.} \& \bibinfo{author}{Yuan, F.}
\newblock \bibinfo{title}{Dynamical structure factors of a two-dimensional
  {Fermi }superfluid within random phase approximation}.
\newblock \emph{\bibinfo{journal}{New Journal of Physics}}
  \textbf{\bibinfo{volume}{22}}, \bibinfo{pages}{093012}
  (\bibinfo{year}{2020}).
\newblock \urlprefix\url{https://doi.org/10.1088/1367-2630/abab3d}.

\bibitem{Zielinski2020}
\bibinfo{author}{Zielinski, T.}, \bibinfo{author}{Ross, B.} \&
  \bibinfo{author}{Gezerlis, A.}
\newblock \bibinfo{title}{Pairing in two-dimensional {Fermi }gases with a
  coordinate-space potential}.
\newblock \emph{\bibinfo{journal}{Phys. Rev. A}}
  \textbf{\bibinfo{volume}{101}}, \bibinfo{pages}{033601}
  (\bibinfo{year}{2020}).
\newblock \urlprefix\url{https://link.aps.org/doi/10.1103/PhysRevA.101.033601}.

\bibitem{Mulkerin2020A}
\bibinfo{author}{Mulkerin, B.~C.}, \bibinfo{author}{Hu, H.} \&
  \bibinfo{author}{Liu, X.-J.}
\newblock \bibinfo{title}{Role of the confinement-induced effective range in
  the thermodynamics of a strongly correlated {Fermi }gas in two dimensions}.
\newblock \emph{\bibinfo{journal}{Phys. Rev. A}}
  \textbf{\bibinfo{volume}{101}}, \bibinfo{pages}{013605}
  (\bibinfo{year}{2020}).
\newblock \urlprefix\url{https://link.aps.org/doi/10.1103/PhysRevA.101.013605}.

\bibitem{Mulkerin2020B}
\bibinfo{author}{Mulkerin, B.~C.}, \bibinfo{author}{Liu, X.-J.} \&
  \bibinfo{author}{Hu, H.}
\newblock \bibinfo{title}{Pseudogap regime of a strongly interacting
  two-dimensional {Fermi }gas with and without confinement-induced effective
  range of interactions}.
\newblock \emph{\bibinfo{journal}{Phys. Rev. A}}
  \textbf{\bibinfo{volume}{102}}, \bibinfo{pages}{013313}
  (\bibinfo{year}{2020}).
\newblock \urlprefix\url{https://link.aps.org/doi/10.1103/PhysRevA.102.013313}.

\bibitem{Wang2020}
\bibinfo{author}{Wang, X.}, \bibinfo{author}{Chen, Q.} \&
  \bibinfo{author}{Levin, K.}
\newblock \bibinfo{title}{Strong pairing in two dimensions: pseudogaps, domes,
  and other implications}.
\newblock \emph{\bibinfo{journal}{New Journal of Physics}}
  \textbf{\bibinfo{volume}{22}}, \bibinfo{pages}{063050}
  (\bibinfo{year}{2020}).
\newblock \urlprefix\url{https://dx.doi.org/10.1088/1367-2630/ab890b}.

\bibitem{He2022}
\bibinfo{author}{He, Y.-Y.}, \bibinfo{author}{Shi, H.} \&
  \bibinfo{author}{Zhang, S.}
\newblock \bibinfo{title}{Precision many-body study of the
  {Berezinskii}-{Kosterlitz}-{Thouless} transition and temperature-dependent
  properties in the two-dimensional {Fermi }gas}.
\newblock \emph{\bibinfo{journal}{Phys. Rev. Lett.}}
  \textbf{\bibinfo{volume}{129}}, \bibinfo{pages}{076403}
  (\bibinfo{year}{2022}).
\newblock
  \urlprefix\url{https://link.aps.org/doi/10.1103/PhysRevLett.129.076403}.

\bibitem{Ramachandran2024}
\bibinfo{author}{Ramachandran, S.}, \bibinfo{author}{Jensen, S.} \&
  \bibinfo{author}{Alhassid, Y.}
\newblock \bibinfo{title}{Pseudogap effects in the strongly correlated regime
  of the two-dimensional {Fermi }gas}.
\newblock \emph{\bibinfo{journal}{Phys. Rev. Lett.}}
  \textbf{\bibinfo{volume}{133}}, \bibinfo{pages}{143405}
  (\bibinfo{year}{2024}).
\newblock
  \urlprefix\url{https://link.aps.org/doi/10.1103/PhysRevLett.133.143405}.

\bibitem{Dean1993}
\bibinfo{author}{Dean, D.}, \bibinfo{author}{Koonin, S.},
  \bibinfo{author}{Lang, G.}, \bibinfo{author}{Ormand, W.} \&
  \bibinfo{author}{Radha, P.}
\newblock \bibinfo{title}{Shell model {Monte Carlo} calculations for
  $^{170}${Dy}}.
\newblock \emph{\bibinfo{journal}{Physics Letters B}}
  \textbf{\bibinfo{volume}{317}}, \bibinfo{pages}{275--280}
  (\bibinfo{year}{1993}).
\newblock
  \urlprefix\url{https://www.sciencedirect.com/science/article/pii/037026939390995T}.

\bibitem{Ormand1994}
\bibinfo{author}{Ormand, W.~E.}, \bibinfo{author}{Dean, D.~J.},
  \bibinfo{author}{Johnson, C.~W.}, \bibinfo{author}{Lang, G.~H.} \&
  \bibinfo{author}{Koonin, S.~E.}
\newblock \bibinfo{title}{Demonstration of the auxiliary-field {Monte Carlo}
  approach for $sd$-shell nuclei}.
\newblock \emph{\bibinfo{journal}{Phys. Rev. C}} \textbf{\bibinfo{volume}{49}},
  \bibinfo{pages}{1422--1427} (\bibinfo{year}{1994}).
\newblock \urlprefix\url{https://link.aps.org/doi/10.1103/PhysRevC.49.1422}.

\bibitem{Wang2017}
\bibinfo{author}{Wang, Z.}, \bibinfo{author}{Assaad, F.~F.} \&
  \bibinfo{author}{Parisen~Toldin, F.}
\newblock \bibinfo{title}{Finite-size effects in canonical and grand-canonical
  quantum {Monte Carlo} simulations for fermions}.
\newblock \emph{\bibinfo{journal}{Phys. Rev. E}} \textbf{\bibinfo{volume}{96}},
  \bibinfo{pages}{042131} (\bibinfo{year}{2017}).
\newblock \urlprefix\url{https://link.aps.org/doi/10.1103/PhysRevE.96.042131}.

\bibitem{Shen2020}
\bibinfo{author}{Shen, T.}, \bibinfo{author}{Liu, Y.}, \bibinfo{author}{Yu, Y.}
  \& \bibinfo{author}{Rubenstein, B.~M.}
\newblock \bibinfo{title}{Finite temperature auxiliary field quantum {Monte
  Carlo} in the canonical ensemble}.
\newblock \emph{\bibinfo{journal}{The Journal of Chemical Physics}}
  \textbf{\bibinfo{volume}{153}}, \bibinfo{pages}{204108}
  (\bibinfo{year}{2020}).
\newblock \urlprefix\url{https://doi.org/10.1063/5.0026606}.

\bibitem{Werner2012}
\bibinfo{author}{Werner, F.} \& \bibinfo{author}{Castin, Y.}
\newblock \bibinfo{title}{General relations for quantum gases in two and three
  dimensions: two-component fermions}.
\newblock \emph{\bibinfo{journal}{Phys. Rev. A}} \textbf{\bibinfo{volume}{86}},
  \bibinfo{pages}{013626} (\bibinfo{year}{2012}).
\newblock \urlprefix\url{https://link.aps.org/doi/10.1103/PhysRevA.86.013626}.

\bibitem{Yang1962}
\bibinfo{author}{Yang, C.~N.}
\newblock \bibinfo{title}{Concept of off-diagonal long-range order and the
  quantum phases of liquid {He} and of superconductors}.
\newblock \emph{\bibinfo{journal}{Rev. Mod. Phys.}}
  \textbf{\bibinfo{volume}{34}}, \bibinfo{pages}{694--704}
  (\bibinfo{year}{1962}).
\newblock \urlprefix\url{https://link.aps.org/doi/10.1103/RevModPhys.34.694}.

\bibitem{Levinsen2015}
\bibinfo{author}{Levinsen, J.} \& \bibinfo{author}{Parish, M.~M.}
\newblock \emph{\bibinfo{title}{Strongly Interacting Two-Dimensional {Fermi}
  Gases}}, \bibinfo{pages}{1--75} (\bibinfo{publisher}{World Scientific},
  \bibinfo{address}{Singapore}, \bibinfo{year}{2015}).
\newblock
  \bibinfo{eprint}{{\href{https://arxiv.org/abs/https://www.worldscientific.com/doi/pdf/10.1142/9789814667746\_0001}{{https://www.worldscientific.com/doi/pdf/10.1142/9789814667746\_0001}}}}.

\bibitem{Nightingale1982}
\bibinfo{author}{Nightingale, P.}
\newblock \bibinfo{title}{Finite?size scaling and phenomenological
  renormalization (invited)}.
\newblock \emph{\bibinfo{journal}{Journal of Applied Physics}}
  \textbf{\bibinfo{volume}{53}}, \bibinfo{pages}{7927--7932}
  (\bibinfo{year}{1982}).
\newblock \urlprefix\url{https://doi.org/10.1063/1.330232}.

\bibitem{Santos1981}
\bibinfo{author}{dos Santos, R.~R.} \& \bibinfo{author}{Sneddon, L.}
\newblock \bibinfo{title}{Finite-size rescaling transformations}.
\newblock \emph{\bibinfo{journal}{Phys. Rev. B}} \textbf{\bibinfo{volume}{23}},
  \bibinfo{pages}{3541--3546} (\bibinfo{year}{1981}).
\newblock \urlprefix\url{https://link.aps.org/doi/10.1103/PhysRevB.23.3541}.

\bibitem{Bertaina2011}
\bibinfo{author}{Bertaina, G.} \& \bibinfo{author}{Giorgini, S.}
\newblock \bibinfo{title}{{BCS-BEC} crossover in a two-dimensional {Fermi
  }gas}.
\newblock \emph{\bibinfo{journal}{Phys. Rev. Lett.}}
  \textbf{\bibinfo{volume}{106}}, \bibinfo{pages}{110403}
  (\bibinfo{year}{2011}).
\newblock
  \urlprefix\url{https://link.aps.org/doi/10.1103/PhysRevLett.106.110403}.

\bibitem{Tan2008}
\bibinfo{author}{Tan, S.}
\newblock \bibinfo{title}{Energetics of a strongly correlated {Fermi }gas}.
\newblock \emph{\bibinfo{journal}{Annals of Physics}}
  \textbf{\bibinfo{volume}{323}}, \bibinfo{pages}{2952--2970}
  (\bibinfo{year}{2008}).
\newblock
  \urlprefix\url{https://www.sciencedirect.com/science/article/pii/S0003491608000456}.

\bibitem{Jensen2020Contact}
\bibinfo{author}{Jensen, S.}, \bibinfo{author}{Gilbreth, C.~N.} \&
  \bibinfo{author}{Alhassid, Y.}
\newblock \bibinfo{title}{Contact in the unitary {Fermi }gas across the
  superfluid phase transition}.
\newblock \emph{\bibinfo{journal}{Phys. Rev. Lett.}}
  \textbf{\bibinfo{volume}{125}}, \bibinfo{pages}{043402}
  (\bibinfo{year}{2020}).
\newblock
  \urlprefix\url{https://link.aps.org/doi/10.1103/PhysRevLett.125.043402}.

\bibitem{Frolich2012}
\bibinfo{author}{Fr\"ohlich, B.} \emph{et~al.}
\newblock \bibinfo{title}{Two-dimensional {Fermi }liquid with attractive
  interactions}.
\newblock \emph{\bibinfo{journal}{Phys. Rev. Lett.}}
  \textbf{\bibinfo{volume}{109}}, \bibinfo{pages}{130403}
  (\bibinfo{year}{2012}).
\newblock
  \urlprefix\url{https://link.aps.org/doi/10.1103/PhysRevLett.109.130403}.

\bibitem{Fenech2016}
\bibinfo{author}{Fenech, K.} \emph{et~al.}
\newblock \bibinfo{title}{Thermodynamics of an attractive {2D} {Fermi }gas}.
\newblock \emph{\bibinfo{journal}{Phys. Rev. Lett.}}
  \textbf{\bibinfo{volume}{116}}, \bibinfo{pages}{045302}
  (\bibinfo{year}{2016}).
\newblock
  \urlprefix\url{https://link.aps.org/doi/10.1103/PhysRevLett.116.045302}.

\bibitem{Fulde1964}
\bibinfo{author}{Fulde, P.} \& \bibinfo{author}{Ferrell, R.~A.}
\newblock \bibinfo{title}{Superconductivity in a strong spin-exchange field}.
\newblock \emph{\bibinfo{journal}{Phys. Rev.}} \textbf{\bibinfo{volume}{135}},
  \bibinfo{pages}{A550--A563} (\bibinfo{year}{1964}).
\newblock \urlprefix\url{https://link.aps.org/doi/10.1103/PhysRev.135.A550}.

\bibitem{Larkin1965}
\bibinfo{author}{Larkin, A.} \& \bibinfo{author}{Ovchinnikov, Y.}
\newblock \bibinfo{title}{Nonuniform state of superconductors}.
\newblock \emph{\bibinfo{journal}{Sov. Phys. JETP}}
  \textbf{\bibinfo{volume}{20}} (\bibinfo{year}{1965}).

\bibitem{Shi2016}
\bibinfo{author}{Shi, H.}, \bibinfo{author}{Rosenberg, P.},
  \bibinfo{author}{Chiesa, S.} \& \bibinfo{author}{Zhang, S.}
\newblock \bibinfo{title}{Rashba spin-orbit coupling, strong interactions, and
  the {BCS-BEC} crossover in the ground state of the two-dimensional {Fermi
  }gas}.
\newblock \emph{\bibinfo{journal}{Phys. Rev. Lett.}}
  \textbf{\bibinfo{volume}{117}}, \bibinfo{pages}{040401}
  (\bibinfo{year}{2016}).
\newblock
  \urlprefix\url{https://link.aps.org/doi/10.1103/PhysRevLett.117.040401}.

\bibitem{Vitali2022}
\bibinfo{author}{Vitali, E.}, \bibinfo{author}{Rosenberg, P.} \&
  \bibinfo{author}{Zhang, S.}
\newblock \bibinfo{title}{Exotic superfluid phases in spin-polarized {Fermi
  }gases in optical lattices}.
\newblock \emph{\bibinfo{journal}{Phys. Rev. Lett.}}
  \textbf{\bibinfo{volume}{128}}, \bibinfo{pages}{203201}
  (\bibinfo{year}{2022}).
\newblock
  \urlprefix\url{https://link.aps.org/doi/10.1103/PhysRevLett.128.203201}.

\bibitem{Rosenberg2017}
\bibinfo{author}{Rosenberg, P.}, \bibinfo{author}{Shi, H.} \&
  \bibinfo{author}{Zhang, S.}
\newblock \bibinfo{title}{Ultracold atoms in a square lattice with spin-orbit
  coupling: charge order, superfluidity, and topological signatures}.
\newblock \emph{\bibinfo{journal}{Phys. Rev. Lett.}}
  \textbf{\bibinfo{volume}{119}}, \bibinfo{pages}{265301}
  (\bibinfo{year}{2017}).
\newblock
  \urlprefix\url{https://link.aps.org/doi/10.1103/PhysRevLett.119.265301}.

\bibitem{Chen2012}
\bibinfo{author}{Chen, G.}, \bibinfo{author}{Gong, M.} \&
  \bibinfo{author}{Zhang, C.}
\newblock \bibinfo{title}{{BCS-BEC} crossover in spin-orbit-coupled
  two-dimensional {Fermi }gases}.
\newblock \emph{\bibinfo{journal}{Phys. Rev. A}} \textbf{\bibinfo{volume}{85}},
  \bibinfo{pages}{013601} (\bibinfo{year}{2012}).
\newblock \urlprefix\url{https://link.aps.org/doi/10.1103/PhysRevA.85.013601}.

\bibitem{Zheng2014}
\bibinfo{author}{Zheng, Z.} \emph{et~al.}
\newblock \bibinfo{title}{{FFLO} superfluids in {2D} spin-orbit coupled {Fermi
  }gases}.
\newblock \emph{\bibinfo{journal}{Scientific Reports}}
  \textbf{\bibinfo{volume}{4}}, \bibinfo{pages}{6535} (\bibinfo{year}{2014}).
\newblock \urlprefix\url{https://doi.org/10.1038/srep06535}.

\end{thebibliography}

\end{document}